\documentclass[aps,prl,reprint,groupedaddress,amsmath,amssymb,aps,floatfix]{revtex4-2}

\usepackage{graphicx}% Include figure files
\usepackage{dcolumn}% Align table columns on decimal point
\usepackage{bm}% bold math
\usepackage{hyperref}
\usepackage{booktabs}
\usepackage{multirow,ragged2e} %
\usepackage{stfloats}
\usepackage{caption, subcaption}
\usepackage{hyperref}
\usepackage{array}
\usepackage[dvipsnames]{xcolor}
\usepackage{siunitx}
\providecommand{\e}[1]{\ensuremath{\times 10^{#1}}}
\begin{document}

\title{Non-Linear Dynamics Induced by Strong Radio-Frequency Fields in ReBCO High Temperature Superconductors}

\author{A. Dhar}
\email{adhar@slac.stanford.edu}
\author{M. E. Schneider}
\author{E. A. Nanni}
\affiliation{SLAC National Accelerator Laboratory, Menlo Park,California,USA}
\author{J. Golm}
\author{P. Krkoti\'{c}}
\author{W. Wuensch}
\author{S. Calatroni}
\affiliation{CERN, Geneva, Switzerland}
\author{N. Lamas}
\author{T. Puig}
\author{J. Gutierrez}
\affiliation{ICMAB-CSIC, Barcelona, Spain}

\date{\today}

\begin{abstract}
Probing the dynamics of superconducting transitions induced by strong electromagnetic fields is vital to designing high power devices leveraging these materials. The development of high temperature superconductors (HTS) is particularly interesting due to critical temperatures ($T_c$) approaching 90\,K, the ability to support high current densities, and their ability to operate in strong static magnetic fields. This work aims to determine the transition dynamics of these materials at radio-frequencies (rf) in the microwave range where they have enormous potential for new applications ranging from particle accelerators to dark matter searches. We have tested two types of coatings formed from rare earth barium copper oxide (REBCO): a film deposited by electron-beam physical vapor deposition, and HTS conductor tapes soldered to a copper substrate with exposed REBCO surfaces. Testing was performed via a hemispherical transverse-electric mode cavity that maximizes the surface rf magnetic field and minimizes the surface electric field on a 2-inch diameter sample. We report on steady-state measurements at low rf power, as well as fully time-resolved transition dynamics on the microsecond timescale seen for the first time with strong surface rf fields.
\end{abstract}
	
\maketitle

\section{Introduction}
Probing the dynamics of transitions in Type-II superconductors is critical to understanding the boundary between the normal conducting state and the superconducting state. Unlike Type-I superconductors, these materials allow some magnetic flux to penetrate them through vortices surrounded by supercurrents~\cite{TypeII}. Pinning of these vortices is what enables high field superconducting magnets to be built for a variety of applications~\cite{wilson_superconducting_2002}. Understanding this boundary is important for any high field applications, including those with alternating magnetic fields like superconducting rf (SRF) cavities.

SRF cavities built with materials such as niobium (Nb) have been instrumental in improving rf efficiency for long pulse and continuous-wave (CW) particle accelerators, but require extremely low temperatures (2-4\,K) for operation well below their $T_c$ of 9\,K ~\cite{LCLSII_LINAC,TESLA,LCLSHE}. This cryogenic operation comes with a wall-plug cooling efficiency of 0.1\%, thus requiring SRF cavities to have exceedingly high quality factors to preserve overall efficiency~\cite{2KPower}. Niobium cavities at 1.3\,GHz have been able to reach quality factors (Q) over $10^{10}$ for an accelerating mode, as compared to room temperature normal conducting rf (NCRF) cavities which typically reach quality factors of $\approx 2-3\times10^4$ for the same mode~\cite{TESLA,CuLBand}.

The development of high temperature superconductors (HTS) is promising due to their significantly higher critical temperatures. Rare earth barium copper oxides (REBCO) are particularly interesting because of their critical temperature of $\sim$90\,K. Furthermore, REBCO is commercially available as coated conductor tapes, with other forms of coatings in constant development, making it suitable for resonant structures. HTS cavities could be cooled by liquid nitrogen, cryocoolers or other simplified cryogenic systems instead of liquid helium, drastically reducing the cryogenic infrastructure required for operation. Structures coated with HTS materials could be used as high Q devices such as linearizers, deflector cells, axion cavities, and pulse compressors~\cite{Golm,akira,Rades}. In addition, HTS materials can maintain superconductivity in high magnetic fields, making them uniquely capable of creating an energy efficient muon cooling channel for a potential muon collider~\cite{muon}.

A key drawback of using any superconductor is the limitation on the induced surface magnetic field, which can cause the structure to rapidly transition to the normal conducting state. In SRF this appears to depend on the superheating field $H_{sh}$\cite{yogi_critical_1977, campisi_high_1985}, which limits the surface magnetic field of niobium structures to approximately 170\,kA/m \cite{TESLA,SRFCav}. This is far less than their NCRF counterparts, which can operate at 500\,kA/m or higher depending on the pulse length~\cite{CbandAPL,VDXband,Cahill2018,tan_demonstration_2022}. When comparing performance in an accelerating mode, SRF cavities are capable of around 50\,MV/m with long pulses, while NCRF cavities are capable of over 250\,MV/m with shorter pulses ($<$1\,\SIUnitSymbolMicro s). Having a new regime which could accommodate higher gradients more readily with \SIUnitSymbolMicro s-scale pulses or longer would open up new applications for particle accelerators as well.

Previous rf studies of REBCO materials suggested that small crystal grains and short coherence lengths gave rise to substantial residual resistance, which limits the surface magnetic field these materials can withstand and, consequently, the accelerating gradient they can generate~\cite{YBCORF,YBCORFLosses}. However, given the recent advances in REBCO structural properties and current carrying capacity~\cite{obradors_coated_2014}, it is vital to reevaluate the performance of REBCO in comparison to copper and niobium for use in rf cavities, in part also due to its promising $H_{sh}$ which is about five times higher than for Nb as it approaches \SI{0}{K}~\cite{claire, powell_nonlinear_1999}.  

We have used rf measurements at 11.424\,GHz to determine the rf surface resistance of REBCO planar samples as a function of temperature, applied surface field, and time. This can be done with a resonant cavity partially coated with the superconductor of interest. This allows for precise measurements of steady-state properties at low rf power, and transient behavior at higher rf power where strong fields are applied to the material surface. We performed measurements on two types of REBCO samples inserted within a hemispherical transverse-electric mode cavity that maximizes the rf magnetic field and minimizes the electric field on a 2-inch diameter sample~\cite{NantistaCav,GuoCav,TScavity}. Based on these methods we aimed to observe any major changes in surface resistance based on proximity to the critical temperature and surface field limits. These observations will allow us to begin to define the parameter space within which HTS materials could be used to build high power rf cavities.

\section{HTS Samples}

\begin{figure}[h]
    \centering
    \begin{subfigure}[h]{.49\columnwidth}
		\caption{}
    \includegraphics*[width=\columnwidth]{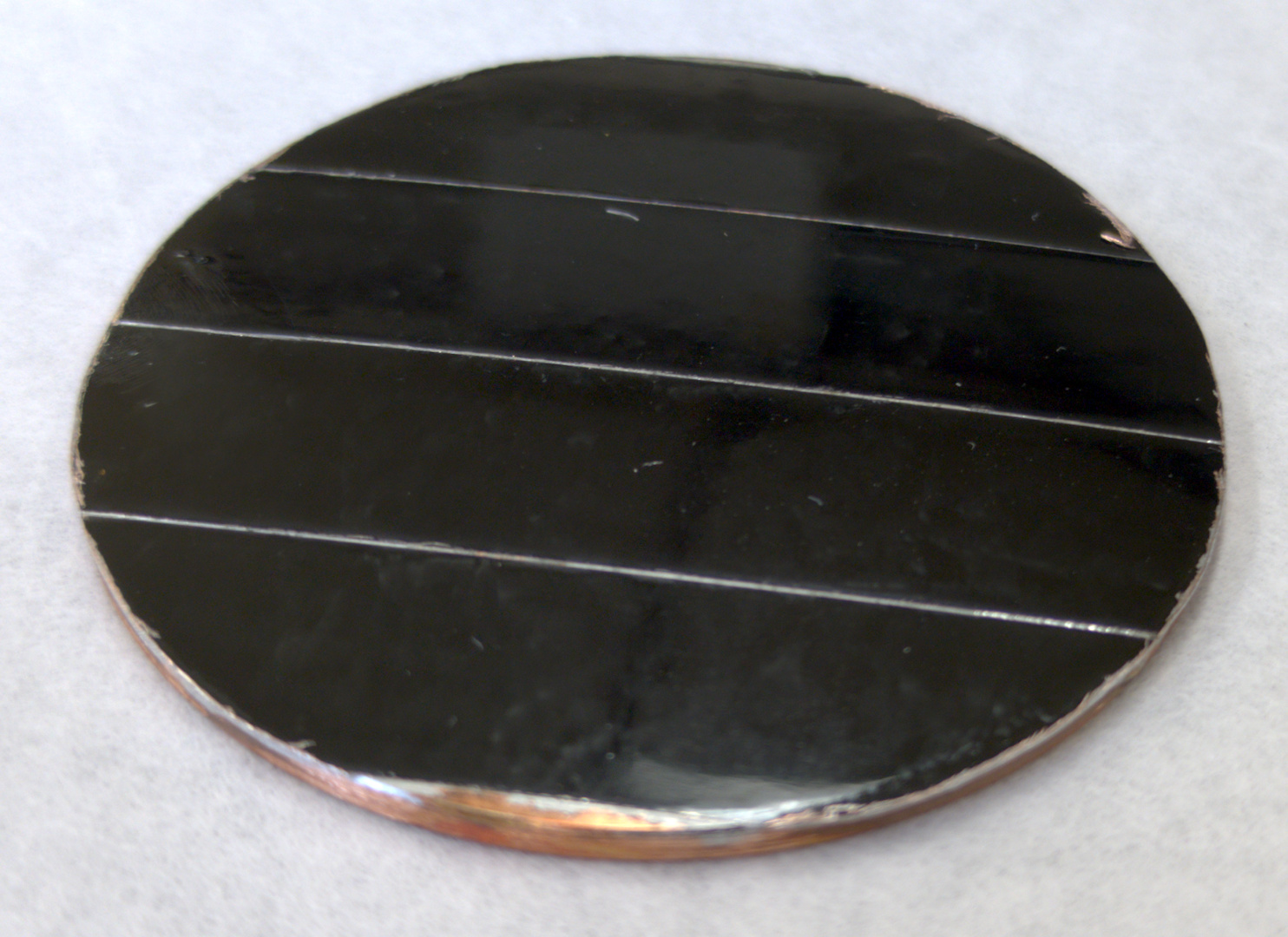}\label{fig:discT}
    \end{subfigure}
    \begin{subfigure}[h]{.49\columnwidth}
		\caption{}
    \includegraphics*[width=\columnwidth,clip=true,trim=0 5 0 7]{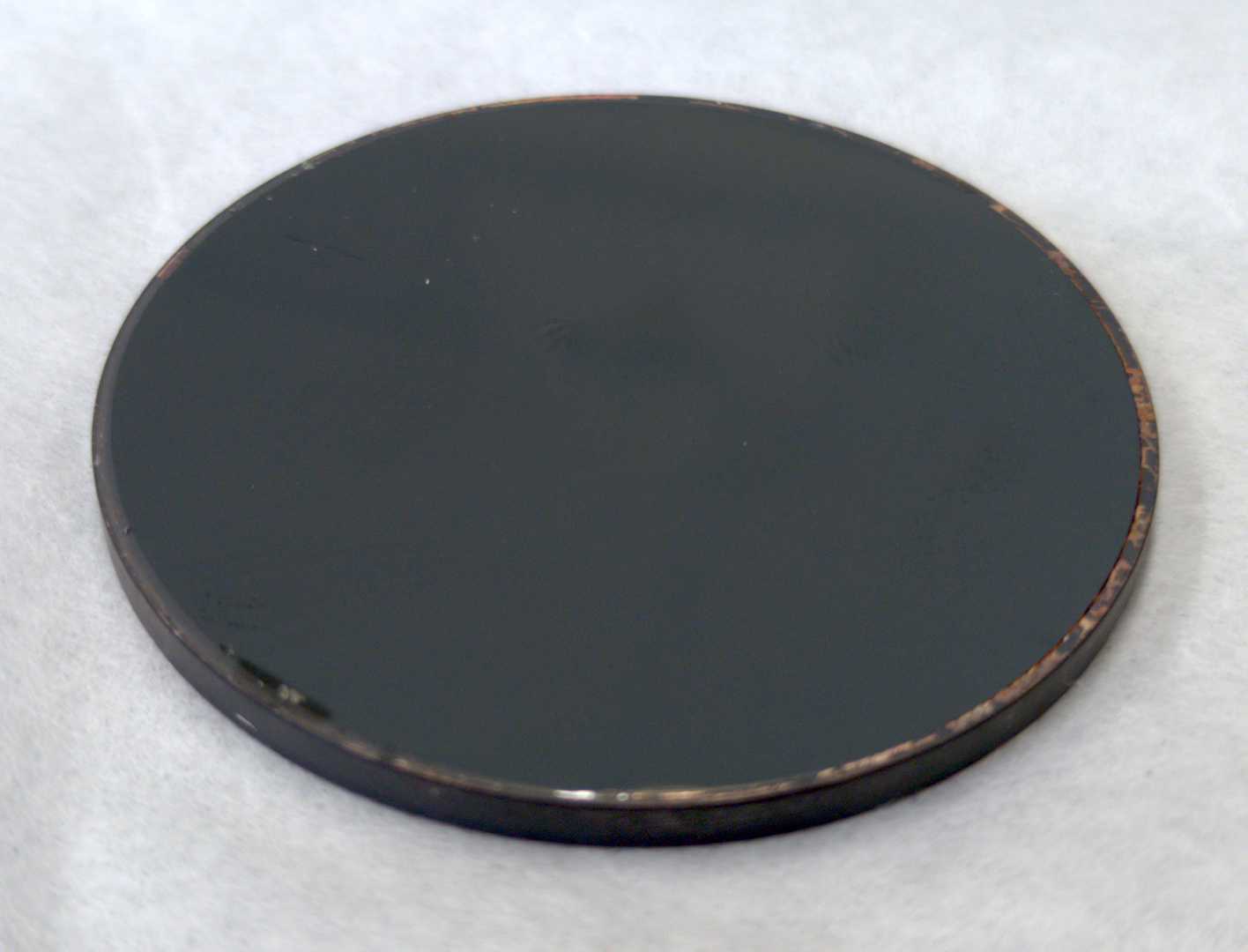}\label{fig:discF}
    \end{subfigure}
    \caption{Photographs of the samples: (a) soldered REBCO-tapes on copper and (b) film directly grown on copper with a MgO buffer layer. Both samples were coated onto 2-inch copper discs.}
    \label{fig:discs}
\end{figure}

Although there are many types of HTS materials, this study focused on REBCO materials specifically. Two distinct types of REBCO samples with different coating techniques and surface treatments were tested. Both samples were adhered to 2-inch diameter copper discs. The first sample, shown in Fig.~\ref{fig:discT}, utilized commercially available coated conductor tapes from Fujikura (FESC-SCH12) made with Europium that were 12 mm wide and included BaHfO$_3$ nanorods as artificial pinning centers~\cite{fujikura}. In this sample, four 12\,mm tapes were initially soldered side-by-side onto a copper disc, followed by delamination between the REBCO and the buffer film on the surface~\cite{Telles}. This approach resulted in the REBCO being the topmost layer, but also introduced normal conducting joints between the superconducting tapes that rf current must flow through. 

For the second sample, the REBCO film was deposited via electron-beam physical vapor deposition and made with Dysprosium, covering the entire surface of the copper disc, as depicted in Fig.~\ref{fig:discF}. A MgO buffer layer was first thermally evaporated and reactively grown on the copper substrate, with a crystal axis tilted by approximately 30$^{\circ}$ from vertical. The REBCO then nucleated on the inclined MgO plane, resulting in the REBCO crystal axis having a similar inclination angle of approximately 30$^{\circ}$\cite{Theva}. This inclination angle results in anisotropic surface resistance in the plane of the film. This would be an important factor to consider when interpreting measurements from within a resonant cavity.

\section{Experimental Setup}
\begin{figure}[h]
   \centering
   \includegraphics*[width=1\columnwidth]{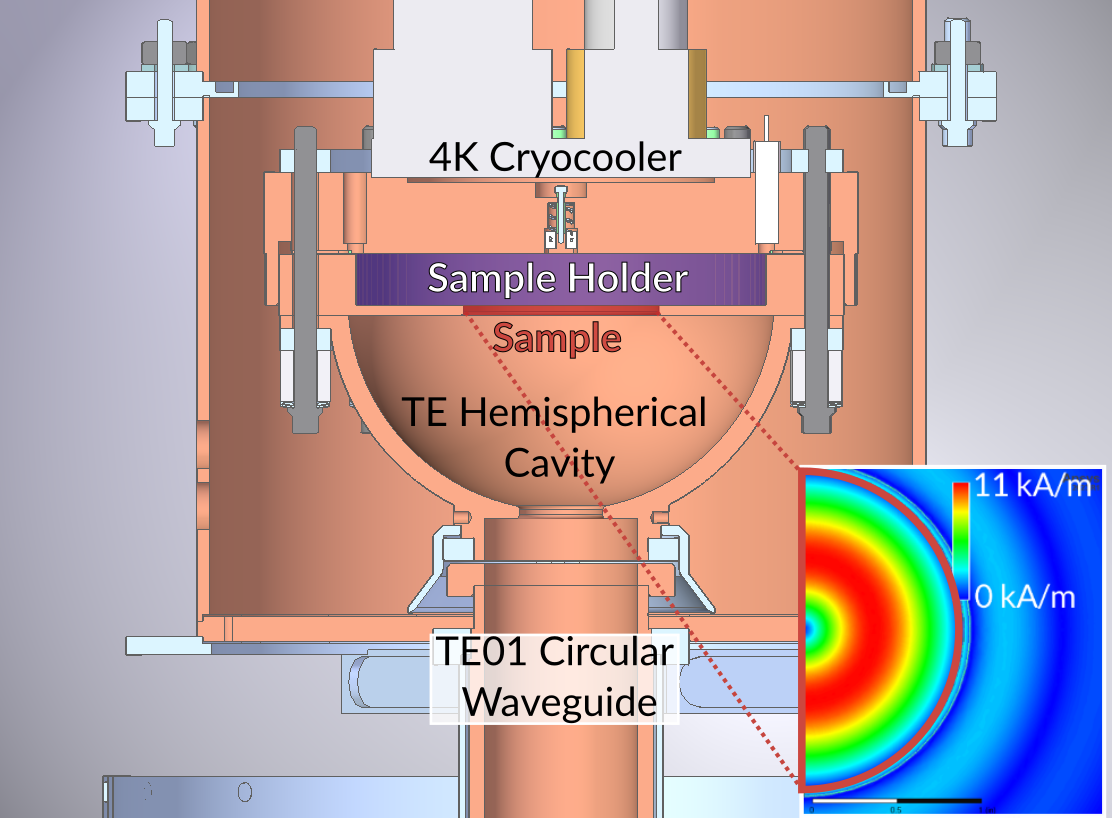}
   \caption{Cross-sectional view of the azimuthally symmetric hemispherical cavity within the cryostat. The sample holder (purple) allows for 2 inch samples (red) to be easily swapped out for rf measurements. The magnetic fields on the sample (insert) have azimuthal symmetry due to the TE$_{023}$ mode within the cavity. Fields are shown for 1.6\,kW of input power.}
   \label{fig:TS4}
\end{figure}

Each sample was mounted within a hemispherical cavity for rf measurements. A schematic of the hemispherical cavity installed in the experimental cryostat can be seen in Fig.~\ref{fig:TS4}. This cryostat can operate down to 4\,K through the use of a pulse tube helium cryo-cooler. The base of the cavity is designed to allow for 2 inch diameter samples to be easily swapped out for rf measurements~\cite{TScavity}. Samples were tested at varying rf power levels within this cavity, which is designed to have a TE$_{023}$ mode centered at 11.424\,GHz~\cite{GuoCav,TScavity}. This mode minimizes the electric field on the sample surface and maximizes the magnetic field. The cavity fields and their frequency dependence are modeled in HFSS~\cite{ANSYS}. This allows us to properly understand any rf measurements taken with this cavity in order to extract sample material properties. 

The hemispherical cavity is coupled into a TE$_{01}$ circular waveguide mode, and can be driven by either a vector network analyzer (VNA) that probes with $<$1\,mW or an X-band traveling wave tube (TWT) that can provide up to 1.6\,kW pulses over 8\,\SIUnitSymbolMicro s. At higher power ($>$100\,W), the cavity is used for time domain measurements of superconducting transition dynamics for a given surface magnetic field. The VNA by comparison allows for steady-state measurements of a given sample's rf surface resistance. 

In both regimes, measurements were conducted by slowly heating the sample up from 4\,K and measuring the reflected power from the cavity. In the steady-state regime, this could be done directly with a Keysight P9373A VNA. For higher power measurements, forward and reflected power were both measured through Keysight N1912A power meters. These power meters were used to measure the pulse shape of forward and reflected power in order to measure the total quality factor of the cavity. Transient information of the sample during these measurements confirmed prior steady-state measurements at low power.

\section{Steady State Measurements}
\begin{figure}[h]
   \centering
   \includegraphics*[width=\columnwidth]{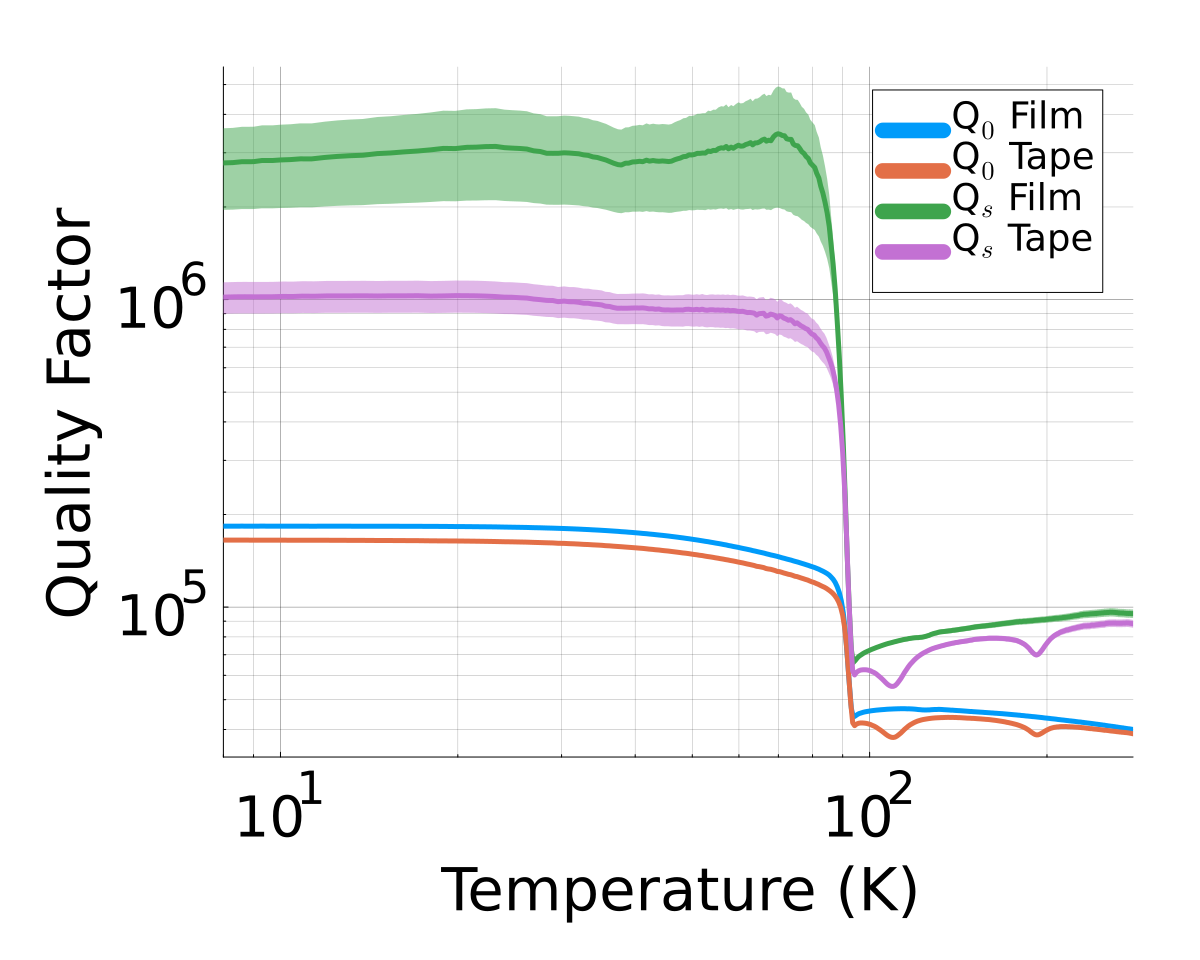}
   \caption{Low power measurements of REBCO tape and film samples, with sample contribution to internal quality factor plotted as well. Shaded regions represent the error in sample quality factor measurement due to uncertainty in the cavity quality factor model. The uncertainty on the REBCO film sample is particularly high because of how close it reaches the model values. The critical temperature for these REBCO materials is 89\,K. Note that the dips in the quality factors above 100\,K are due to the cavity mode passing through and being superimposed upon a waveguide mode.}
   \label{fig:REBCO}
\end{figure}

The intrinsic sample rf surface resistance as a function of temperature was determined with low-power steady-state measurements. These measurements assess the effective surface resistance of the sample which would account for any damage or imperfections. The sample quality factor ($Q_{s}$) can be obtained from the internal quality factor ($Q_{0}$), which is a combination of $Q_{s}$ and the quality factor of the cavity itself ($Q_\mathrm{cav}$),
\begin{equation}\label{Q0}
        \frac{1}{Q_0}=\frac{1}{Q_s} + \frac{1}{Q_\mathrm{cav}}.
\end{equation}
Here $Q_{0}$ is derived from a circle fit of the complex reflection coefficient ($S_{11}$), while $Q_\mathrm{cav}$ represents the loss from all cavity walls except the sample region. By measuring the cavity with copper (which has the same conductivity as the walls) and niobium (which has a surface resistance far lower than the walls), it is possible to build an empirical model for $Q_\mathrm{cav}$ (see Appendix). Using this model and Eq.~\ref{Q0}, the sample quality factor can be extracted, as shown in Fig.~\ref{fig:REBCO}.

From here, $Q_{s}$ can be used to derive an equivalent rf surface resistance of the sample ($R_{s}$) by using a reference conductivity for bulk Cu $\sigma_\mathrm{ref}$ and then calculating a reference $Q_{s\mathrm{,ref}}$ for the cavity with a bulk Cu sample in it. This measurement provides a relative comparison between the samples being studied, and does not represent a direct measure of BCS resistance or residual resistance.
%Based on the fact that Q $\propto$ $\sqrt{\sigma}$, 
The equivalent sample surface resistance is given as

\begin{align}\label{sigs}
        &Q =G/R \text{, where } R_s=\sqrt{\frac{\mu\omega}{2\sigma}}\nonumber\\
    \therefore\, &\sigma_s=\sigma_\mathrm{ref}\left(\frac{Q_s}{Q_{s\mathrm{,ref}}}\right)^2,\nonumber\\
    \therefore\, &R_{s,sample}=\sqrt{\frac{\mu\omega}{2\sigma_s}}.
\end{align}

The reference conductivity for copper is 58\,MS/m at room temperature, which would give a $Q_{s\mathrm{,ref}}$ of 1.55$\times10^5$~\cite{NIST}. This ratio was then used to derive the quantities shown in Table~\ref{REBCOTable}. At 4\,K, the equivalent rf sample resistance of the taped sample was approximately twice than that of the film sample, partially due to normal conducting gaps between each tape on the taped sample. Regardless, the equivalent surface resistance of both REBCO samples was significantly lower than that of copper, but still higher than niobium at these frequencies. These measurements form the foundation necessary to probe these materials with even stronger fields and higher rf power. 

\begin{table}[h]
   \centering
   \caption{Summary of steady-state measurements at 4\,K and 80\,K for copper and niobium as compared to REBCO tape and film samples.}
    \begin{tabular}{m{.175\columnwidth}m{.2\columnwidth}m{.25\columnwidth}m{.25\columnwidth}}
       \toprule
       \textbf{Sample} & \hfil\textbf{$Q_0$}& \hfil\textbf{$Q_s$} & \hfil\textbf{$R_s$ [m$\Omega$]} \\
       \midrule
       \multicolumn{4}{c} {4\,K}\\
       \midrule
           Niobium & \hfil{1.93$\e{5}$} & \hfil{-}  & \hfil{-}\\ 
           HTS Film & \hfil{1.85$\e{5}$} &  \hfil{3.3 $\pm$ 1.1$\e{6}$} & \hfil{1.45 $\pm$ 0.5}      \\
           HTS Tape & \hfil{1.71$\e{5}$} &  \hfil{1.2 $\pm$ 0.2$\e{6}$} & \hfil{3.60 $\pm$ 0.5}      \\ 
            Copper     & \hfil {1.40$\e{5}$} &  \hfil{4.8 $\pm$ 0.3$\e{5}$} & \hfil{9.02 $\pm$ 0.5}      \\ 
           \midrule
           \multicolumn{4}{c} {80\,K}\\
           \midrule
           %REBCO Film & \hfil{1.4$\e{5}$} & \hfil{1$\e{6}$} & \hfil{2.6}      \\ 
           HTS Film & \hfil{1.35$\e{5}$} & \hfil{2.5 $\pm$ 0.9  $\e{6}$} & \hfil{2.03 $\pm$ 0.7}      \\ 
           HTS Tape & \hfil{1.21$\e{5}$} & \hfil{7.4 $\pm$ 0.9$\e{5}$} & \hfil{5.72 $\pm$ 0.4}      \\
           Copper     & \hfil{1.01$\e{5}$} & \hfil{3.4 $\pm$ 0.2$\e{5}$} & \hfil{12.3 $\pm$ 0.6}      \\ 
           \midrule
           \multicolumn{4}{c} {300\,K}\\
           \midrule
        Copper & \hfil{4.90$\e{4}$} & \hfil{1.55$\e{5}$} & \hfil{27.91}   \\ 
       \bottomrule
  \end{tabular}
  \label{REBCOTable}
\end{table}

\section{Transition dynamics due to strong fields near T$_c$}
With the steady-state performance well characterized, the next step was to measure transition dynamics by applying stronger surface fields to the sample. These measurements were performed with input rf power ranging from 100\,W to 1.6\,kW, which corresponds to 2.5 to 14\,mT peak surface magnetic field on the sample. Each pulse was 8\,\SIUnitSymbolMicro s long, with a repetition rate of 100\,Hz. Forward power was ramped up slowly at each temperature setpoint, in order to better ascertain the relationship between sample equivalent rf surface resistance and applied surface field near the superconducting transition point.

\begin{figure}[h]
	\centering
    \includegraphics[width=\columnwidth]{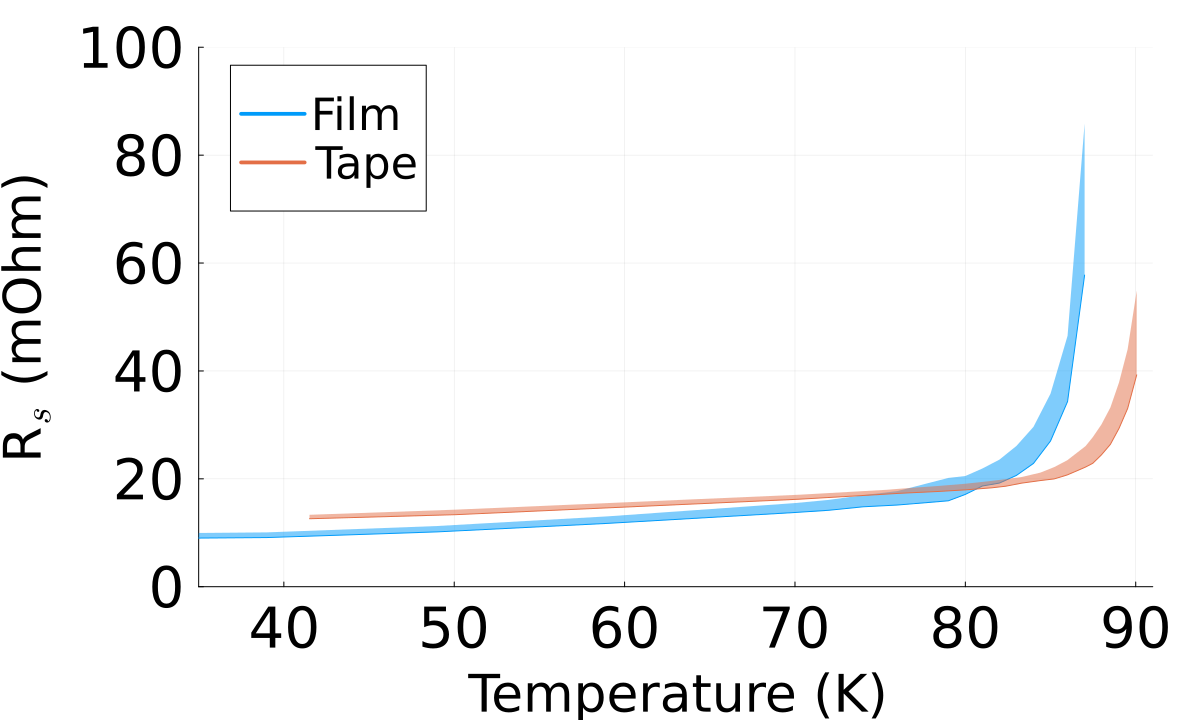}
	\caption{Surface resistance plot for both REBCO samples at the end of the rf pulse. The shaded regions represent the range of values as a function of forward power, between 1600\,W (higher surface resistance) and 100\,W (lower surface resistance).}
	\label{fig:sigma}
\end{figure}

The quality factor of the cavity after each pulse was determined by looking at the decay in reflected power. The time constant $\tau$ of the reflected power's exponential decay is related to the cavity's total quality factor based on $Q_t=2\pi f_0\tau$, where $f_0$ is the resonant frequency of the cavity in MHz. This time constant is determined by plotting the reflected power on a semi-log plot and fitting a line to the data, resulting in $\tau=-\frac{1}{m}$, where $m$ is the slope of slope of the semi-log decay in reflected power. Prior analysis of the steady-state data determined that the external quality factor ($Q_{e}=1.4\times10^5$) was roughly constant over the temperature range of interest (4\,K to 100\,K). This meant that $Q_{0}$ could be determined from the following equation:
\begin{equation}
        \frac{1}{Q_0}=\frac{1}{Q_{t}} - \frac{1}{Q_{e}}.
\end{equation}
From here the analysis is similar to the low power measurements, extracting the contribution of the sample from the internal quality factor and in turn its equivalent rf surface resistance as shown in Eqs.~\ref{Q0} and \ref{sigs}. These results are summarized in Fig.~\ref{fig:sigma}. While the film sample was able to reach a lower equivalent rf surface resistance that is closer to the steady-state value, it also was more drastically affected by applied fields and began transitioning closer to 86\,K instead of the expected critical temperature of 89\,K. This is likely due to the anisotropic surface resistance of the film, forcing some regions with lower critical current limits to transition at lower temperatures than expected. In comparison, the taped sample shows greater variation in surface resistance with respect to applied power, but does not exhibit a dramatically large rise until 89\,K. This could be driven by the gaps between the tapes, which would form hot spots that then translate to greater rise in overall surface resistance. These measurements were still limited to observing the state of the sample at the end of the rf pulse. The next logical step was to observe how surface resistance evolves during the rf pulse.

\begin{figure}[h]
	\centering
    \begin{subfigure}[h]{\columnwidth}
		\caption{}
    \includegraphics*[width=1\columnwidth]{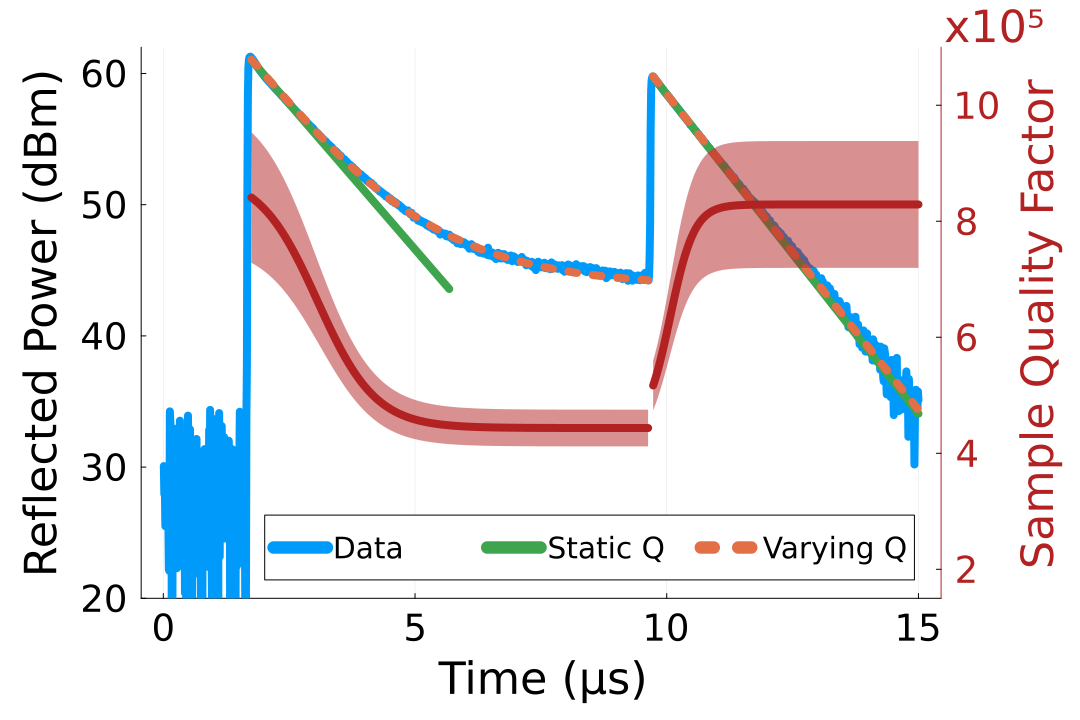}\label{fig:QofT}
    \end{subfigure}
    \begin{subfigure}[h]{\columnwidth}
		\caption{}
    \includegraphics*[width=1\columnwidth]{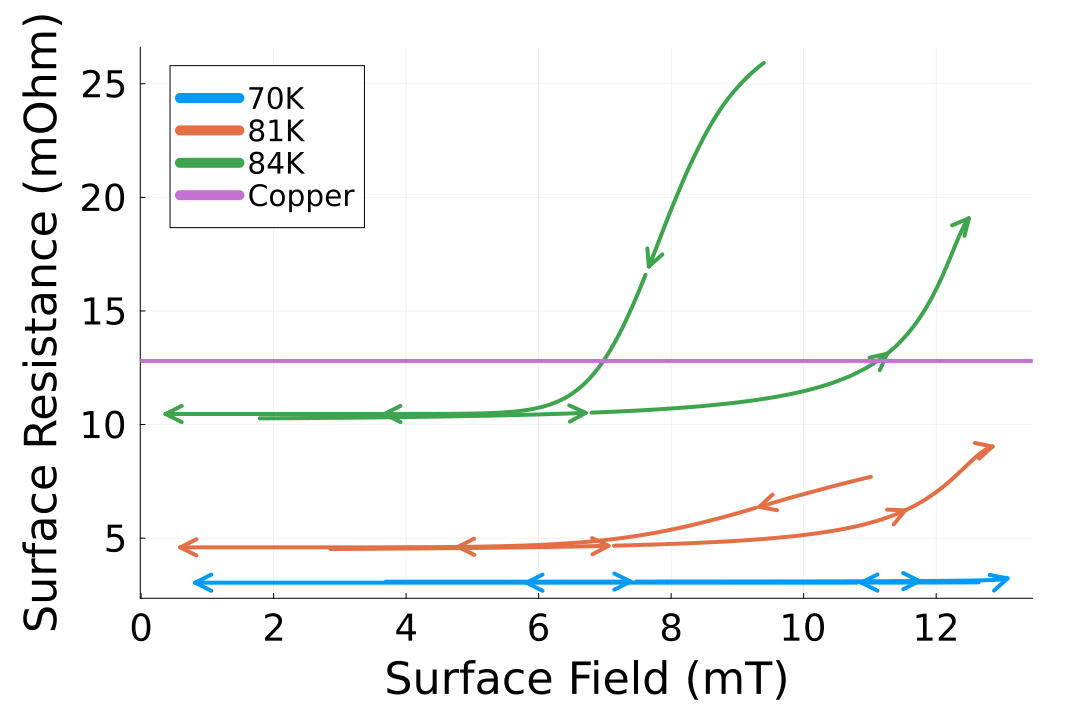}\label{fig:SofT}
    \end{subfigure}
	\caption{(a) Plot of reflected power (blue) from cavity with the film sample at 81\,K for 62\,dBm input power. At this power level, a constant quality factor model (green) no longer fits this decay accurately. Using a time-varying Q model (orange) fits the data correctly, determining Q$_s$ as a function of time (red). (b) This time-varying sample quality factor can then be used to compare the surface resistance of the sample at high surface magnetic field at varying temperatures (blue, orange,green). Approaching  85\,K (the observed high power transition point for the film sample), the sample surface resistance rises to the point where it exceeds the surface resistance of copper (purple), suggesting that a significant portion of the sample has transitioned to the normal conducting state. The larger the overall transition, the more time required for the sample to recover its superconductivity by the next pulse.}
	\label{fig:quench}
\end{figure}

Figure~\ref{fig:QofT} shows an example of a reflected power measurement for the film sample represented in dBm. The decay in power (shown in blue) during the rf pulse represents the state of the cavity as it fills with rf energy, while the decay after the pulse represents energy flowing out of the cavity. Looking now at the decay during the rf pulse, initial attempts at fitting the data to a constant quality factor were not successful (shown in green). To correct for this, a time-varying model of the internal quality factor (shown in red) was used to correctly fit the data (shown in orange)~\cite{Cahill2018}. This revealed that the initial quality factor starts at a higher value closer to the steady state, before rapidly dropping over several microseconds. Similarly, it quickly recovers after the rf pulse, implying that transition to the normal conducting state was driven by rf fields within the cavity, as opposed to residual heat warming the sample past the critical point. 

Due to the anisotropic conductivity of the film sample, and the normal conducting gaps between tapes, it remains an open question as to which specific regions on the sample are transitioning. This also does not account for any hot spots that may occur from surface impurities or areas of bad electrical or thermal conductivity~\cite{hotspot}. This will be addressed in upcoming tests of samples with isotropic conductivity, and new test cavities which focus the surface magnetic field along only one axis. However, based on models for pulsed surface heating from high power rf pulses, the expected peak temperature rise within the sample should be no more than 0.3\,K\cite{pulsed}. 

Instead, the sample is likely transitioning due to the applied surface field pushing the material past a critical current density, which would explain why it recovers as fields decay within the cavity. This is made more clear by comparing the surface resistance of the sample to the expected surface magnetic fields within the cavity, shown in Fig.~\ref{fig:SofT}. Here it is more apparent that the increase in resistance at higher field is correlated how close to the critical temperature the sample is. For example,at 70\,K the surface resistance remain less than \SI{10}{\milli\ohm} and is not strongly affected by surface field. Contrasting this, at 81\,K and more so at 84\,K the surface resistance sharply rises with surface field, and slowly recovers after in a hysteric fashion. This confirms that the sample is able to recover its superconductivity before the next rf pulse. 

Thus far we have tested REBCO high temperature superconductors up to 14\,mT peak surface magnetic field, which would generate an accelerating gradient of roughly 3\,MV/m in a standard L-band SRF cavity~\cite{TESLA}. We found that these fields can cause the material dynamically transition around 81\,K, very close to the critical temperature. Based on how quickly the surface resistance changes during and after the rf pulse, our measurement approach has allowed us to distinguish between transitions driven by surface fields and those driven by surface heating.

\section{Conclusion}
Understanding these limits and mechanisms behind them will require observing these same transitions at lower temperatures with substantially higher rf power. This will require MW-scale rf power sources like klystrons, which can provide higher power at the cost of pulse length, which in turn may make observation of time-dependent phenomena within the pulse more difficult to discern. This work represents the initial efforts which are being built upon to test the limits of HTS at lower temperatures ($<$80\,K) and higher field strengths ($>$95\,mT) which would correspond to 20\,MV/m in a standard L-band SRF cavity. Together, all of these studies will be crucial to determining the feasibility of HTS cavities made from REBCO materials for high power rf applications.

% Specify following sections are appendices. Use \appendix* if there
% only one appendix.
\appendix*
\section{Quality Factor of Cavity Derivation}
In order to characterize the sample testing cavity's contribution to losses, we built a model of its contribution to the internal quality factor ($Q_\mathrm{cav}(T)$). To do this, we tested a niobium and copper sample in the cavity. When the niobium sample is superconducting, we can treat it as a perfect conductor setting a lower bound of 
\begin{align}
  Q_\mathrm{cav}(4\,K) \approx Q_{0}(4\,K),
\end{align}
since $Q_{s\mathrm{,Nb}} >> Q_\mathrm{cav}$. Based on these measurements, this means $Q_{\mathrm{cav}}(4\,K)$ would be 1.9375$\times10^5$.

With the copper sample in place, the entire cavity is now formed from copper. This means that the functional form of $Q_{0}(T)$ with copper should be directly proportional to $Q_\mathrm{cav}$, given as 
\begin{align}
  Q_\mathrm{cav}(T) &= C\times Q_{0\mathrm{,Cu}}(T),
\end{align}
where $C = Q_{\mathrm{cav}}(4\,K)$. Fitting $Q_{0\mathrm{,Cu}}(T)$ to a spline and combining with the Nb results gives us a model for $Q_\mathrm{cav}(T)$, which was used to extract the effective surface resistance of any sample. Comparing these results for copper to known values of conductivity for copper at varying Residual Resistance Ratios (RRR) allowed us to set an upper bound for $C$ of 2.02$\times10^5$. This allowed us to define the uncertainty in our effective surface resistance measurements, which increases as $Q_{s} \rightarrow Q_\mathrm{cav}$.

% If you have acknowledgments, this puts in the proper section head.
\begin{acknowledgments}
The authors would like to thank Valery Dolgashev, Valery Borzenets, Greg Le Sage, Matt Boyce, Paul Welander, Pablo Martinez, Sami Tantawi and Ruggero Vaglio for many helpful discussions. This work is supported by U.S. Department of Energy Contract No. DE-AC02-76SF00515. 
\end{acknowledgments}

% Create the reference section using BibTeX:
\bibliography{biblio.bib}

\end{document}